\documentstyle[12pt]{article}
\newcommand{\nc}{\newcommand}
\def\frac#1#2{{\textstyle {#1 \over #2}}}

\nc{\beq}{\begin{equation}}
\nc{\eeq}{\end{equation}}
\nc{\beqa}{\begin{eqnarray}}
\nc{\eeqa}{\end{eqnarray}}
\nc{\lsim}{\begin{array}{c}\,\sim\vspace{-21pt}\\< \end{array}}
\nc{\gsim}{\begin{array}{c}\sim\vspace{-21pt}\\> \end{array}}
\def\ZZ{\hbox{\it Z\hskip -4.pt Z}}

\def\RR{\hbox{\it I\hskip -2.pt R }}
\def\tvi{\vrule height 12pt depth 6pt width 0pt}
\def\tv{\tvi\vrule}
\def\cc#1{\kern .7em\hfill #1 \hfill\kern .7em}

\begin{document}
\begin{titlepage}

\begin{center}
\vskip .5 in
{\large \bf Non-Trivial Extensions of the $3D-$Poincar\'e Algebra}\\ 
{\large \bf and Fractional 
Supersymmetry for Anyons}
\vskip .3 in
{
  {\bf M. Rausch de Traubenberg}\footnote{rausch@lpt1.u-strasbg.fr}
   \vskip 0.3 cm
   {\it  Laboratoire de Physique Th\'eorique, Universit\'e Louis Pasteur}\\
   {\it 3-5 rue de l'universit\'e, 67084 Strasbourg Cedex, France}\\ 
  \vskip 0.3 cm
and \\
{\bf M. J. Slupinski\footnote{slupins@math.u-strasbg.fr}}\\
{\it Institut de Recherches en Math\'ematique Avanc\'ee}\\
{ \it Universit\'e Louis-Pasteur, and CNRS}\\
{\it 7 rue R. Descartes, 67084 Strasbourg Cedex, France}\\ } 
\end{center}

\vskip .5 in
\begin{abstract}
Non-trivial extensions of the three dimensional Poincar\'e algebra, beyond 
the supersymmetric one, are explicitly  constructed. These algebraic 
structures are the natural three dimensional generalizations of fractional 
supersymmetry of order $F$ already considered in one and two dimensions.
Representations of these algebras are exhibited, and 
unitarity  is explicitly checked.
It is then  shown that
these extensions generate symmetries which  connect fractional spin states
or anyons. Finally, a natural classification arises according to the 
decomposition of $F$ into its product of prime numbers leading to
sub-systems
with smaller symmetries.

\end{abstract}
\end{titlepage}

\renewcommand{\thepage}{\arabic{page}}
\setcounter{page}{1}

In $D-$dimensional spaces,  particles are classified by irreducible 
representations of the Poincar\'e algebra. This  algebra generates the
space-time symmetries (Lorentz transformations and space-time
translations), 
and after one has
gauged the space-time translations  we naturally obtain  a theory of
gravity.
Therefore, in order to understand the fundamental interactions and the
symmetries
in particle physics, it is interesting to study all the possible extensions
of 
the Poincar\'e symmetry. Quantum Field Theory restricts considerably the 
possible
generalizations. If one imposes the unitarity of the $S-$matrix with a
discrete
spectrum of massive one particle states, then
within the framework of Lie algebras,
the Coleman and Mandula theorem \cite{cm}  allows only
 internal symmetries, {\it i.e.} those 
commuting with the generators of the Poincar\'e algebra\footnote{
In the massless case, the Poincar\'e group can be promoted to the conformal
one.}. However, if we go beyond Lie algebras, we can escape this 
{\it no-go} theorem. The   well-known supersymmetric extension is generated
by
fermionic charges which, by the Haag, Lopuszanski and Sohnius 
theorem,
are in the spinorial representation of $SO(1,D-1)$ \cite{hls}. 
So, it seems that there exists
 a {\it unique} non-trivial extension of the Poincar\'e algebra, up
to the choice of the number $N$ of supercharges. Indeed, according to the
Noether theorem, all these symmetries correspond to conserved currents, and
are generated by charges which are expressed in terms of the fields.
By
the spin-statistics theorem we have two kinds of fields having integer or
half-integer spin. The former will close with commutators and the latter
with
anticommutators leading respectively to Lie and super-Lie algebras.

The consideration of algebraic extensions, beyond the Poincar\'e algebra,
is not new. Such a possibility was considered in \cite{ker, luis}. 
In the second paper, Wills Toro showed  that the generators of the
Poincar\'e
algebra might themselves have non-trivial  indices. In this paper
we pursue a different possibility, namely the study of special 
dimensions. Particular dimensions can reveal 
exceptional behaviour. This opportunity to find ``particular'' dimensions
has already  been  exploited with success and has 
led to generalizations of
supersymmetry. Fractional supersymmetry (FSUSY) which was introduced  
in  \cite{fsusy}, is one such  generalization.
In one-dimensional spaces, where
no rotation is available, this symmetry is generated by one 
generator which can be seen as the $F^{th}$ root of the time translation
$\left(Q_t\right)^F=\partial_t$. $F=2$ corresponds to the usual
supersymmetry.
A group theoretical justification was then given in \cite{am,fr}
and this symmetry was applied in the world-line formalism  \cite{fr}. 
The second peculiar cases, are the  two-dimensional
spaces where, by  use of conformal transformations, the 
(anti)holomorphic part  of the fields transforms independently \cite{bpz}.  
In \cite{prs}, this situation
was exploited to build a Conformal Field Theory with fractional conformal
weight. The Virasoro algebra was extended by two generators satisfying 
$\left(Q_z\right)^F=\partial_z$ and $\left(Q_{\bar z}\right)^F=
\partial_{\bar z}$  and besides the stress-energy tensor,  a conserved
current of conformal weight ($1 + {1 \over F}$) was obtained. 
Several groups have also studied this symmetry
in one \cite{fsusy1d} and two dimensions \cite{fsusy2d}.

Finally, in $1+2$ dimensions particles with arbitrary spin and statistics
exist. The so-called anyons were defined  for the first time
in \cite{lm}. In fact, studying the representations of the $3D-$ Poincar\'e
algebra $P_{1,2}$ the unitary irreducible representations divide into two 
classes: massive or massless.
For the massive particles, we can consider a one-dimensional wave function
with arbitrary real spin$-s$ ({\it i.e.} which picks up an arbitrary phase 
factor $\exp(2i\pi s)$ when rotated through $2 \pi$).  
In the massless case, only
two types of  discrete spin exist \cite{b}. Then a relativistic wave
equation
for anyons was formulated following different approaches in \cite{jn,p}.

The purpose of this letter  is to build non-trivial extensions of the
Poincar\'e 
algebra which go beyond supersymmetry (SUSY).
We first give a fractional supersymmetric extension of
the Poincar\'e algebra of any order $F$.  Then, 
we study the representations of this algebra which turn out to contain
anyonic fields with spin ($\lambda,\lambda  -{1 \over F}, \cdots,
\lambda  -{F-1 \over F}$) (in the simplest case and with  $\lambda$ an
arbitrary real number). We also explicitly check that the representations
we are considering are unitary. 

It then appears that $3D-$FSUSY, like in $2D$, is a symmetry
which connects the fractional spin states previously obtained. 
In this sense it is a natural 
generalization of SUSY. We also prove that the algebras so-obtained can be
classified according to the decomposition of $F$ into its product
of prime numbers. 

Introducing the generators of space-time translations $P^\alpha$ and the 
generators of Lorentz 
transformations \-$J^\alpha = {1 \over 2} 
\eta^{\alpha \beta}$ $ \epsilon_{\beta \gamma \delta} J^{\gamma \delta}$, 
 we can rewrite the three dimensional Poincar\'e algebra as
follows

\beqa
\label{eq:P}
\left[ P^\alpha ,P^\beta \right]  &=& 0 \nonumber \\
\left[ J^\alpha ,P^\beta \right] &=& i \eta^{\alpha \gamma} \eta^{\beta
\delta}
\epsilon_{\gamma \delta \eta} P^\eta \\
\left[ J^\alpha,J^\beta \right] &=&  i \eta^{\alpha \gamma} \eta^{\beta
\delta}
\epsilon_{\gamma \delta \eta} J^\eta,    \nonumber
\eeqa

\noindent
with $\eta_{\alpha \beta} = {\mathrm{diag}}(1,-1,-1)$  the Minkowski metric
and  $\epsilon_{\beta \gamma \delta}$
the completely antisymmetric Levi-Civita tensor such that 
$\epsilon_{012} =1$. 
\noindent
Particles are then classified according to the values of the Casimir
operators of the Poincar\'e algebra. More precisely, for  a mass $m$
particle
of positive/negative energy, 
the unitary irreducible representations are obtained by studying the little
group leaving the rest-frame momentum $P^\alpha=(m,0,0)$ invariant. This
stability group in $\overline{SO(1,2)}$, the universal covering group of
${SO(1,2)}$,
is simply the universal covering group $\RR$ of 
the abelian sub-group of rotation  $SO(2)$ (generated by $J^0$).
As it is well-known, such a group is not quantized. This means that the 
substitution $J^0 \to J^0 + s$ leaves the $SO(2)$ part invariant. But the 
remarkable property of $\overline{SO(1,2)}$, is that the concomitant transformation
on the Lorentz boosts $J^i \to J^i + s { P^i \over P^0 + m}$  leaves the
algebraic structure (\ref{eq:P}) unchanged. Anyway, following the method
of induced representation for groups expressible as  a semi-direct product
we 
find that unitary irreducible representations for a massive particles are
one dimensional, and that the Lorentz generators are \cite{jn,b}
(for an arbitrary spin$-s$ representation)

\beqa
J^0_s &=& i \left(p^1 {\partial \over \partial p_2}
 - p^2 {\partial \over \partial p_1}\right)  
+ s \nonumber \\
J^1_s &=& - i \left(p^2 {\partial \over \partial p_0}- 
p^0 {\partial \over \partial p_1}\right) 
+ s { p^1 \over p^0+m} \\
J^2_s &=& - i \left(p^0 {\partial \over \partial p_1}- 
p^1 {\partial \over \partial p_0}\right)
+ s { p^2 \over p^0+m}, \nonumber
\eeqa

\noindent
with $p^\alpha$ the eigenvalues of the operators $P^\alpha$. This
modification
of the $3D$ Lorentz generators was pointed out in \cite{sch} and is not
the most general one we can consider (see the last paper of \cite{p}).

\noindent
The main difference between   $SO(1,2)$, or more precisely the
proper orthochronous Lorentz group, and $SO(3)$ is that  $p^0+m$ never
vanishes with $SO(1,2)$ and $s$ does not need  to be quantized.

In Ref.\cite{jn,p},  a relativistic wave equation for
massive anyons was given.
First,  notice that the two Casimir operators are the
two scalars $P.P$ and $P.J$ and their eigenvalues for a spin$-s$ unitary
irreducible representation are respectively $m^2$ and $ms$. The  equations
of
motion are then

\beqa
\label{eq:ms}
(P^2 - m^2) \Psi &=& 0 \\
(P.J - sm) \Psi  &=& 0.  \nonumber
\eeqa

However, to obtain  manifestly covariant equations one has to go beyond the
mass-shell conditions (\ref{eq:ms}) given by the induced representation. 
Therefore, we can start with a field which belongs to the appropriate
spin$-s$
representation of the {\it full} Lorentz group instead of the little group.
When $s$ is a negative  integer, or a negative half-integer, this 
representation is not unitary and is  $2|s|+1$
dimensional, and the solution of the relativistic wave equations reduces to
the appropriate induced representation (see \cite{b,jn} for an
explicit calculation in the case $|s|=1,1/2$). When $s$ is an arbitrary
number,
the representation is infinite dimensional and belongs to the discrete
series of $\overline{SO(1,2)}$ \cite{wy}. A relativistic wave equation for an anyon
in the continuous series \cite{wy} was also considered in the
third paper of \cite{p}.  
Noting $J_{s,\pm}=J^1_s \mp i~J^2_s$ 
($[J^0,J_\pm]=\pm J_\pm,~~[J_+,J_-]=-2J^0$)
the Lorentz generators of the spin$-s$
representation, and $|s,n \rangle$ the states ($n =0,\dots, \infty$)
we can build two spin$-s$ representations;
one bounded from below, noted ${\cal D}^+_s$

\beqa
\label{eq:sb}
J_s^0 |s_+,n \rangle &=& (s+n) |s_+,n \rangle \nonumber \\
J_{s,+} |s_+,n \rangle &=& \sqrt{(2s+n)(n+1)} |s_+,n+1 \rangle \\
J_{s,-} |s_+,n \rangle &=& \sqrt{(2s+n-1)n} |s_+,n-1 \rangle, \nonumber
\eeqa
\noindent
and one bounded from above (${\cal D}^-_s$)
\beqa
\label{eq:sa}
J_s^0 |s_-,n \rangle &=& -(s+n) |s_-,n \rangle  \nonumber \\
J_{s,+} |s_-,n \rangle &=& - \sqrt{(2s+n-1)n}|s_-,n-1 \rangle  \\
J_{s,-} |s_-,n \rangle &=& - \sqrt{(2s+n)(n+1)} |s_-,n+1 \rangle. \nonumber
\eeqa

\noindent
For both representations, the quadratic Casimir operator
of the Lorentz group equals $s(s-1)$.
For the first representation we have $J_{s,-} |s_+,0 \rangle =0$ and 
for the second  $J_{s,+} |s_-,0 \rangle =0$. 
Jackiw and Nair \cite{jn} and Plyushchay \cite{p} 
were able to define an equation of motion (plus some
subsidiary conditions) such that the solution of a spin$-s$ anyonic
equation
decomposes into a direct sum of a positive energy solution in the 
representation bounded from
below and a negative energy in the one bounded from above. In
other
words, a solution of a spin$-s$ anyonic equation decomposes into
a positive energy state of helicity $h=s$ and a negative energy solution
with $h=-s$ : $|s \rangle = |h=s, +\rangle \oplus |h=-s,- \rangle$ and the
two states are $CP$ conjugate.

If $s$ is a negative integer or a negative half-integer  number we get a 
$2|s|+1$  dimensional representation,
but for a general $s$ we have an infinite number of states. Furthermore when
$s < 0$ the representation is non-unitary. Taking the spinorial representation
as a guidline, we choose the case $s=-1/F$ to build a non-trivial extension
of the Poincar\'e algebra. If we observe the  relations (\ref{eq:sa}) and 
(\ref{eq:sb}) with $s=-1/F$, we see an ambiguity in the square
root of $-2/F$. So a priori we have four different representations for 
$s=-1/F$, (two bounded from below/above) with the two choices 
$\sqrt{-1}=\pm i$.
We note ${\cal D}^\pm_{-1/F;\pm}$ (with obvious notations) these 
representations. Next, we can
make the following identifications
\begin{itemize}
\item the dual representation of  ${\cal D}^+_{-1/F;+}$ is obtained through
the substitution $J^a \longrightarrow -\left(J^a\right)^t$ and is given by
$\left[{\cal D}^+_{-1/F;+}\right]^*={\cal D}^-_{-1/F;+}$;
\item the complex conjugate representation of ${\cal D}^+_{-1/F;+}$ is
defined by $J^a \longrightarrow -\left(J^a\right)^\star$ \footnote{In the
mathematical literature because in the definition of Lie algebras there is no
$i$ factor --see equation (\ref{eq:P})-- we do not have a minus sign
in the definition of this representation.} 
\footnote{Note that, for a complex matrix $X$, $X^\star$ denotes the
complex conjugate (and not the hermitian conjugate) matrix of $X$;
for a vector space $V$, $V^*$ is its dual.}
 (we have to be careful when we do such
a transformation because we have {\it by definition $J^\pm = J^1 \mp i J^2$,
for any representation})
is given by $\overline{{\cal D}^+_{-1/F;+}}={\cal D}^-_{-1/F;-}$;
\item the dual of the complex conjugate representation of 
${\cal D}^+_{-1/F;+}$ is given by 
$\left[\overline{{\cal D}^+_{-1/F;+}}\right]^*={\cal D}^+_{-1/F;-}$.
\end{itemize}
If we note $\psi_a \in {\cal D}^+_{-1/F;+} ,\psi^a \in {\cal D}^-_{-1/F;+},
\bar{\psi}_{\dot a} \in {\cal D}^-_{-1/F;-}$ and 
$\bar{\psi^{\dot a}} \in {\cal D}^+_{-1/F;-}$  then we  have the following
transformation laws:

\beqa
\psi^\prime_a &=& S_a^{~~b} \psi_b \nonumber \\
\psi^{\prime a} &=& \left(S^{-1}\right)_{b}^{~~a} \psi^b  \\
\bar{\psi}^\prime_{\dot a} &=& \left(S^\star\right)_{\dot a}^{~~ \dot b} 
\bar {\psi}_{\dot b} 
\nonumber \\
\bar{\psi}^{\prime {\dot a}}&=& \left((S^\star)^{-1} \right)_{\dot b}^
{~~ \dot a}  \bar{\psi}^{\dot b}. \nonumber
\eeqa 

\noindent
Furthermore, if we define 
\beq
\psi^a = g^{a \dot a} \bar \psi_{\dot a},
\eeq
\noindent
we can write the following scalar product

\beq
\label{eq:ps}
\varphi^a \psi_a = -\bar \varphi_{\dot 0} \psi_0 + \sum \limits_{a>0} 
\bar \varphi_{\dot a} \psi_a,
\eeq 
\noindent
where the infinite matrix $g^{a \dot a}$ and its inverse $g_{\dot a a}$
are given by ${\mathrm{diag}} (-1,1,\cdots,1)$.
The reason why we have a pseudo-hermitian scalar product is because
we are dealing with a non-unitary representation of a
non-compact Lie group. The invariant scalar product gives an explicit
isomorphism between the two representations bounded from below (or above)
($\left(S^{-1}\right)_b^{~~a}= g^{a \dot a} 
\left( S^\star\right)_{\dot a}^{~~\dot b} g_{\dot b b}$).

From now on, we choose  $\sqrt{-2/F}=i\sqrt{2/F}$ for 
representations bounded from below and  $\sqrt{-2/F}=-i\sqrt{2/F}$ for 
those bounded from above.

Using the representations (\ref{eq:sb}--\ref{eq:sa}),
and with the sign ambiguity resolved, we can define two
series
of operators, belonging to a non-trivial representation of the Poincar\'e
algebra. We denote now $\sqrt{-1}=i$. 
Note $Q^+_{-1/F+n}$ those built from the representation bounded from below
(${\cal D}^+_{-1/F;+}$)
and $Q^-_{-1/F+n}$  the charges of the representation bounded from above
(${\cal D}^-_{-1/F;-}$).
Using (\ref{eq:sa}, \ref{eq:sb}) we get

\beqa
\label{eq:Q}
\left[J^0,Q^+_{-1/F+n} \right] &=& (n-1/F)~~ Q^+_{-1/F+n}  \nonumber \\
\left[J_+,Q^+_{-1/F+n} \right] &=& \sqrt{(-2/F+n)(n+1)}~~ Q^+_{-1/F+n+1} 
\nonumber
\\
\left[J_-,Q^+_{-1/F+n} \right] &=& \sqrt{(-2/F+n-1)n}~~ Q^+_{-1/F+n-1} 
\nonumber \\
&& \\
\left[J^0,Q^-_{-1/F+n} \right] &=& - (n-1/F) ~~Q^-_{-1/F+n}  \nonumber \\
\left[J_+,Q^-_{-1/F+n} \right] &=& - \left(\sqrt{(-2/F+n-1)n}\right)^\star
~~ Q^-_{-1/F+n-1} 
\nonumber
\\
\left[J_-,Q^-_{-1/F+n} \right] &=& - \left(\sqrt{(-2/F+n)(n+1)}\right)^\star
~~Q^-_{-1/F+n+1}.
\nonumber
\eeqa

\noindent
  
We want to combine this algebra (\ref{eq:Q}) in a non-trivial way with the
Poincar\'e algebra (\ref{eq:P}). With such a choice, $Q^+_{-1 \over F}$ (resp.
$Q^-_{-1 \over F}$) has a helicity $h=-{1\over F}$ (${1 \over F}$ resp.).
With the above choices for the square roots of the negative numbers we know
that the representations are conjugate to each other {\it i.e.} 
$\left(Q_{-1/F+n}^+\right)^\dag \equiv Q_{-1/F+n}^-$.

Having set the values of $s$, we have two reasons 
to close the algebra with
the $Q$'s through a $F^{th}-$order product. First of all, we would like 
the
algebra to be a direct generalization of the one built in two-dimensions. 
Second,
the charges  we have introduced are in the spin$-{1 \over F}$
representation 
of the  Poincar\'e algebra, and so the $Q$'s pick up an 
$\exp{(-{2i\pi \over F})}$ phase factor when rotated through
$2\pi$. They have a non-trivial $\ZZ_F$  graduation, although
the generators of the Poincar\'e algebra are trivial with
respect to $\ZZ_F$. The algebra splits then into an 
anyonic  $\cal{A}$ and a bosonic $\cal{B}$ part. It can be written 

\beqa
\label{eq:PQ0}
&&\left\{\cal{A},\cdots, \cal{A} \right\}_F \sim \cal{B} \nonumber \\
&&\left[\cal{B},\cal{A}\right]  \sim \cal{A} \\
&&\left[\cal{B},\cal{B}\right]  \sim \cal{B}, \nonumber
\eeqa
\noindent
with $\{{\cal A}_{s_1}, \cdots,{\cal A}_{s_F} \}_F={1 \over F !} 
\sum\limits_{\sigma \in \Sigma_F}
{\cal A}_{i_{s_{\sigma(1)}}} \cdots {\cal A}_{i_{s_{\sigma(F)}}}$ and
$\Sigma_F$ 
the  permutation group with $F$ elements.   
Equations (\ref{eq:PQ0}) reveal the $\ZZ_F$ structure  of the algebraic 
extension of the Poincar\'e algebra we are considering. The bosonic part
of the algebra is generated by $J$ and $P$ and has a graduation zero.
The anyonic generators are the supercharges $Q^\pm$ and have graduation $\mp 1$
in $\ZZ_F$. To close the algebra, both sides of the equation have to have
the same graduation, justifying (\ref{eq:PQ0}). In the case of the
supersymmetric extension of the Poincar\'e algebra,  (\ref{eq:PQ0})
corresponds to a $\ZZ_2-$graded Lie algebra or a superalgebra.

Now, we want to identify the whole algebraic extension of $P_{1,2}$.
Part of this algebra is known (see eqs.(\ref{eq:P}) and (\ref{eq:Q})).
Using  adapted Jacobi   identities, we calculate the remaining part of the
algebra, and justify the use of a completely symmetric product in 
(\ref{eq:PQ0}). Those involving three bosonic fields
or two bosonic and one anyonic fields are the same as for 
superalgebras. Using the Leibniz rule of ${\cal B}$ with $\{\dots\}_F$ 
we get the third Jacobi identity and the last one is obtained by a 
direct calculation

\beqa
\label{eq:J}
&&\left[\left[{\cal B}_1,{\cal B}_2\right],{\cal B}_3\right] + 
\left[\left[{\cal B}_2,{\cal B}_3\right],{\cal B}_1\right] +
\left[\left[{\cal B}_3,{\cal B}_1\right],{\cal B}_2\right] =0 \nonumber \\
&&\left[\left[{\cal B}_1,{\cal B}_2\right],{\cal A}_3\right] +
\left[\left[{\cal B}_2,{\cal A}_3\right],{\cal B}_1\right] +
\left[\left[{\cal A}_3,{\cal B}_1\right],{\cal B}_2\right]  =0 \nonumber \\
&&\left[{\cal B},\left\{{\cal A}_1,\dots,{\cal A}_F\right\}_F\right] =
\left\{\left[{\cal B},{\cal A}_1 \right],\dots,{\cal A}_F\right\}_F  +
\dots +
\left\{{\cal A}_1,\dots,\left[{\cal B},{\cal A}_F\right] \right\}_F \\
&&\sum\limits_{i=1}^{F+1} \left[ {\cal A}_i,\left\{{\cal A}_1,\dots,
{\cal A}_{i-1},
{\cal A}_{i+1},\dots,{\cal A}_{F+1}\right\}_F \right] =0. \nonumber
\eeqa

In order to identify the whole algebraic structure of the non-trivial 
extension of the Poincar\'e algebra, assume, as a first step,  
$\left[\-|\cal{A},\cdots, \cal{A}|\-\right]_F = \alpha. P + \beta. J,$

\noindent
with $[| \cdots |]$ a symmetric product of charges to be defined.
If we use the  third Jacobi identity with ${\cal B} =P$, we obtain
$\beta =0$ ( $\left[ P, Q \right] =0$), the same Jacobi identity with
$ {\cal B} = J^0$ proves that both sides of the equation 
have the same helicity.
In other words, this equation just tells us that we need to build
a mapping from  a {\bf sub-space} of ${\cal S}^F({\cal D}_{-{1 \over F}}^\pm)$ 
(the $F-$fold symmetric product of
the representation ${\cal D}_{-{1 \over F}}^\pm$) to
the vectorial ($P$) representation of 
$SO(1,2)$ which is equivariant for the action of $SO(1,2)$. 

Now, we remark that there are primitive states in 
${\cal S}^F({\cal D}_{-{1 \over F}}^\pm)$ from which we are 
able to construct the vector representation of $SO(1,2)$:

\beqa
\label{eq:vrep}
\left[J^0,\left(Q^{\pm}_{-{1 \over F}}\right)^F\right]&=& 
\mp \left(Q^{\pm}_{-{1 \over F}}\right)^F \\
\left[J_\mp,\left(Q^{\pm}_{-{1 \over F}}\right)^F\right]&=&0 \nonumber
\eeqa 
 
\noindent
From these relations, it follows that  the sub-space

$${\cal D}_{-1}=\left\{\left(Q^{\pm}_{-{1 \over F}}\right)^F,
\left[J_\pm,\left(Q^{\pm}_{-{1 \over F}}\right)^F\right],
\left[J_\pm,\left[J_\pm,\left(Q^{\pm}_{-{1 \over F}}\right)^F\right]\right] \right\}$$
 
\noindent
of ${\cal S}^F({\cal D}_{-{1 \over F}}^\pm)$ is
isomorphic to the vector representation of the Poincar\'e algebra.
Note that this relations also imply that
$\left[J_\pm,\left[J_\pm,\left[J_\pm, \left(Q^\pm_{-{1 \over F}}\right)^F
\right]\right]\right] = 0$. \\

So, we obtain the following algebra
(we have to be careful with the normalization appearing in the bracket
$\left\{\cdots \right\}$, for instance $\left(Q^\pm_{-{1 \over
F}}\right)^{F-1}
Q^\pm_{1-{1 \over F}} + \cdots Q^\pm_{1-{1 \over F}} 
\left(Q^\pm_{-{1 \over F}}\right)^{F-1} = F \left\{Q^\pm_{-{1 \over F}},
\cdots,Q^\pm_{-{1 \over F}}, Q^\pm_{1-{1 \over F}}\right\}$).
 
\beqa
\label{eq:PQ}
&&\left\{Q^\pm_{-{1\over F}},\dots,Q^\pm_{-{1\over F}} \right\}_F = P_\mp
\nonumber \\
&&\left\{Q^\pm_{-{1\over F}},\dots,Q^\pm_{-{1\over F}},Q^\pm_{1-{1\over F}}
\right\}_F
 =\pm i \sqrt{{2 \over F}}  P^0 \\
&&   -(F-1) 
\left\{Q^\pm_{-{1\over F}},\dots,Q^\pm_{-{1\over F}},Q^\pm_{1-{1\over F}},
Q^\pm_{1-{1\over F}} \right\}_F
\pm i \sqrt{ F-2}
\left\{Q^\pm_{-{1\over F}},\dots,Q^\pm_{-{1\over F}},Q^\pm_{2-{1\over F}} 
\right\}_F
=   P_\pm \nonumber \\
&&\left[J_\pm,\left[J_\pm,\left[J_\pm, \left(Q^\pm_{-{1 \over F}}\right)^F
\right]\right]\right]=0 \nonumber \\
&& ~~~~~~~~~~~~ \vdots \nonumber
\eeqa

\noindent
with $P_\pm = P^1 \mp i P^2$. The normalization of the R.H.S. of
eq.(\ref{eq:PQ}) comes from the definition of the bracket 
$\left\{\cdots\right\}_F$ and  (\ref{eq:P},\ref{eq:Q}).
Now, we can address the question of the remaining brackets? 
In fact, it is impossible  to find a decomposition\footnote{We thank the
referee for pointing this tu us.}

\beq
{\cal S}^F\left({\cal D}^\pm_{-1/F}\right) = {\cal D}_{-1} \oplus V,
\eeq

\noindent
where $V$ is stable under $SO(1,2)$. Indeed, if there were such a
decomposition there would be a $SO(1,2)$ equivariant projection

\beq
\pi:~{\cal S}^F\left({\cal D}^\pm_{-1/F}\right) \longrightarrow {\cal D}_{-1}.
\eeq 

\noindent  
But then 
$X^\pm=\pi\left( {\cal S}^F\left(Q^\pm_{-1/F},\cdots,Q^\pm_{-1/F},Q^\pm_{3-1/F}
\right) \right) \in {\cal D}_{-1}$ satisfies (see \ref{eq:Q})

$$\big[J_\mp,\big[J_\mp,\big[J_\mp,X^\pm \big]\big]\big]= 
\pm i\sqrt{2/F}\sqrt{2(1-2/F)} \sqrt{3(2-2/F)} P_-\ne 0,$$

\noindent
and this is impossible because in the vector representation ${\cal D}_{-1}$,
$J_-^3$ acts as zero.

Finally, we can note that  direct calculation easily shows that equations 
(\ref{eq:PQ})
are stable under hermitian conjugation. 

In this family of algebras, noted $FSP_{1,2}$ if
we take $F=2$ we are in an exceptional situation.
First, instead of having an infinite number of charges we have only two.
Secondly, the two representations $Q^\pm$ are equivalent
whereas  the two series of charges are inequivalent
representations of $SO(1,2)$ when $F\neq 2$. In the case $F=2$, with
one series of supercharges $Q$ we obtain the well-known supersymmetric
extension of the Poincar\'e algebra, and (\ref{eq:Q}), (\ref{eq:PQ})  can
be
easily rewritten with the Pauli matrices. For more details on this algebra,
one can see, for example, the book of Wess and Bagger \cite{wb}.
The algebra we have obtained is then a direct generalization of the 
super-Poincar\'e one.
It is remarkable  that the supersymmetric algebra, which can
be generalized easily in one and two-dimensional spaces, can also
be considered in $1+2$ dimensions. This is a consequence of the special
feature of $SO(1,2)$ which allows  to define states with
fractional statistics, {\it i.e.} anyons. If we try to go beyond, and
to build an extension of SUSY for higher dimensional spaces, one immediately
 faces an obstruction. Indeed, when $D \ge 4$ one just has
bosonic or fermionic states and supersymmetry is the unique non-trivial
extension of the Poincar\'e algebra one can build.

Finally, let us mention that, the similarity of the algebra 
(\ref{eq:PQ}) and the SUSY algebra
does not stop at this point.
If one considers now $N$  series of charges $Q^+$ and $Q^-$
we obtain, as in SUSY, algebraic extensions with central charges.

Before studying the representations of the algebra 
(\ref{eq:PQ}) we can  address
some general properties. First, $P^2$ commutes with all the generators so
that   all states in an irreducible representation have the same mass.
Secondly, if
we define an anyonic-number operator $\exp({2i\pi{\cal N}_A})$ which gives
the 
phase $e^{2i\pi s}$ on a spin$-s$ anyon we have ${\mathrm{tr}} 
\exp({2i\pi{\cal N}_A}) =0$  showing that
in each irreducible representation there are $F$ possible statistics
($s,s-{1\over F},\dots, s-{ {F-1\over F}}$, where $s$  will be specified
later) and the dimension of the space with a given statistics is
always
the same.  This can be checked proving by that ($\exp({2i\pi{\cal N}_A})
Q_s = e^{2i\pi s}  Q_s \exp({2i\pi{\cal N}_A})$) and using cyclicity of
the trace


\beqa
&&{\mathrm{tr}} \left( \exp({2i\pi{\cal N}_A})
\left\{Q^+_{-{1\over F}},\dots, Q^+_{-{1\over F}},Q^+_{1-{1\over
F}}\right\}_F\right) \nonumber \\
&=&1/F \times {\mathrm{tr}}\left( \sum \limits_{a=0}^{F-1} e^{2i\pi{\cal N}_A} 
\left(Q^+_{-{1\over F}}
\right)^a
\left(Q^+_{1-{1\over F}}\right)\left(Q^+_{-{1\over F}}\right)^{F-a-1}\right)
\nonumber \\
&=& 1/F \times \left(\sum \limits_{a=0}^{F-1} e^{-{2i \pi a \over F}}\right)
{\mathrm{tr}} \left(
\left(Q^+_{-{1\over F}}\right)^{F-1}e^{2i\pi{\cal N}_A}
 \left(Q^+_{1-{1\over F}}\right)\right)=0.
\nonumber 
\eeqa

\noindent
Of course because we are dealing with infinite dimensional algebras
the construction of the trace should be done with care.
However, we will explicitly see, by constructing the unitary representations,
that  ${\mathrm{tr}} \exp({2i\pi{\cal N}_A}) =0$. 

Having defined the anyonic  extensions of the Poincar\'e algebra,
we now look at the massive representations of  (\ref{eq:P}),
(\ref{eq:Q}) and (\ref{eq:PQ}).
Up to now we have written the algebra in such a way that there
is still one ambiguity: we do not know whether we can  choose an
algebraic extension of the Poincar\'e algebra using only one series of
supercharges ($Q^+$ or $Q^-$) or whether we need both. In fact the unitarity 
of the  representation will force us to take {\it both simultaneously}.
Let us first concentrate on the case where one
series of supercharges is involved, say $Q^+$.   
For the Poincar\'e as well as for its
supersymmetric extension, the irreducible representations are obtained,
using the Wigner method of induced representation. Then,
the massive representations $p^\alpha p_\alpha=m^2$  are constructed by
studying the sub-algebra leaving the rest-momentum  $p^\alpha=(m,0,0)$
invariant. Similarly, within the framework of the FSUSY algebras,
the one particle-states are characterised by the eigenvalue
of the rotation in the $(x^1,x^2)$ plane {\it i.e.} by the helicity.
In other words, all the representations are obtained by studying the
sub-algebra where $P_\pm,J_\pm$ are set to zero.  
On the level of the charges,
a similar assumption will be made (valid {\it only } on shell):
if we are looking at eqs.(\ref{eq:PQ}) only one fundamental bracket does
not vanish, 
{\it i.e} the one involving $(F-1)$ times the charge $Q_{-{1 \over F}}$ and
the one involving $Q_{1-{1 \over F}}$ once. All  brackets involving the 
$Q_{n-{1 \over F}}$'s with $n>1$ acts trivially on the rest-frame states
(the R.H.S. always vanishes), so those charges can be represented by $0$
(this is not a new feature and this already appears in usual SUSY, and
for instance in four dimensions, in the massless case,
 the surcharges $Q_2$ and $Q_{\dot 2}$ vanish)).
After an appropriate normalization  (\ref{eq:PQ}) becomes
 
\beqa
\label{eq:PQr}
&&\left\{Q^+_{-{1 \over F}}, \dots,Q^+_{-{1 \over F}},Q^+_{1-{1 \over F}}
\right\}_F
 = 1/F  \\
&&\left\{Q^+_{s_1}, \dots,Q^+_{s_F} \right\}_F = 0,~~ ,
i_1,\cdots i_F=-1/F,1-1/F~~ \mathrm {and },~~i_1 + \dots + i_F \ne 0.
\nonumber
\eeqa 

\noindent
Let us stress  some properties of the algebras defined by (\ref{eq:PQr}).
This kind of algebra is known to mathematicians, and
is called the Clifford algebra of the polynomial $x^{F-1}y$ \cite{rr}. 
Indeed, using  (\ref{eq:PQr})
we obtain (developing explicitly the $F^{th}$ power) $\left(xQ_{-{1\over
F}} +
yQ_{1-{1\over F}}\right)^F=x^{F-1}y$. Hence, the algebra generated by the
two
charges $Q_{-{1\over F}}$ and $Q_{1-{1\over F}}$ is associated with the 
linearization of the polynomial $x^{F-1}y$ and constitute a generalization
of
usual Clifford algebras. This procedure can be considered for any 
polynomial.
However, this algebra  does not
admit a finite  dimensional faithful representation. 
This means that, using a faithful
representation, we are able to build representations with an infinite
number 
of states. It was shown in \cite{frr} that the Clifford algebra of a
polynomial of degree greater than $2$
admits a non-trivial, finite but not faithful representation. 
For $F=2$, the situation is slightly different because  Clifford algebras
admit a finite dimensional faithful representation in terms of the Dirac
$\gamma-$matrices.
Because we want to have a representation which contains a finite number
of states, we consider non-faithful representations.

An extensive study of the representations of Clifford algebras of cubic 
polynomials was undertaken by Revoy \cite{re} and a family of 
representations can be obtained. This result can be generalized for
$F \ge 4$. To obtain the irreducible representations for arbitrary  $F$ we
first observe that $F$ is the first power of $Q_{-{1\over F}}$ which
is equal to zero (in other words the rank of $Q_{-{1\over F}}$  is $F-1$).
Indeed, if one assumes  $Q^{F-b}_{-{1\over F}}=0$ (with $b > 1$), and 
multiplies
the first equation of (\ref{eq:PQr}) by $Q_{-{1\over F}}$ on the left and 
$Q^{F-b-2}_{-{1\over F}}$ on the right, one gets a contradiction.
Using the Jordan decomposition and the property that all  eigenvalues of
$Q_{-{1\over F}}$ are zero, we can write 
\beq
\label{eq:q}
Q^+_{-{1\over F}}=\pmatrix{0&0&0&\ldots&0&0& \cr
                         1&0&0&\ldots&0&0& \cr
                         0&1&0&\ldots&0&0& \cr
                         &\cr
                         \vdots&\vdots&&\ddots&\ddots&\vdots   \cr
                         0&0&\ldots&0&1&0& }
~~\mathrm{and}~~ Q^+_{1-{1\over F}}=
                  \pmatrix{0&0&0&\ldots&0&1 \cr
                          0&0&0&\ldots&0&0& \cr
                          0&0&0&\ldots&0&0& \cr
                          &\cr\
                          \vdots&\vdots&&&\ddots&\vdots& \cr 
                          0&0&0&\ldots&0&0&}.
\eeq
The matrix  representation of $Q_{1-{1\over F}}$ has been obtained, 
solving  (\ref{eq:PQr}). 
When $F=3$ the matrix given in (\ref{eq:q}) for $Q_{1-{1\over F}}$
is not the only possibility \cite{re}, and probably other 
representations can be obtained when $F \ge 4$ \footnote{ This property
was pointed out to us by Ph. Revoy.}. However, the matrices given in
(\ref{eq:q})  are the only ones 
consistent with the Poincar\'e
algebra: if some of the matrix elements which are 
equal to zero in (\ref{eq:q})  are different from zero, 
we obtain equations where both sides do
not have the same helicity (see  below).
Finally, using  the property
that the dimensions of the representations of Clifford algebras
are a multiple of the degree of the polynomial \cite{ht} ($F$ in this
case), 
by similar arguments we can prove that
the other representations are reducible and are built with the two
matrices given in (\ref{eq:q}).

However, the matrices exhibited are not convenient to prove that the 
representations of the FSUSY algebra are unitary. Indeed, we need quadratic
relations upon the matrices $Q^+$ and $Q^-=\left(Q^+\right)^\dag$. So, instead
of the two $Q$'s given on (\ref{eq:q})
(and their hermitien conjugate matrices) we would prefer
more suitable matrices obtained after a rescaling. At least two interesting
solutions have been found (the second was suggested by the referee)

{\tiny
\beqa
\label{eq:q1}
\begin{array}{ll}
Q^+_{-{1\over F}}=\left\{ 
\begin{array}{l}\pmatrix{0&0&0&\ldots&0&0& \cr
                         \sqrt{[1]}&0&0&\ldots&0&0& \cr
                         0&\sqrt{[2]} &0&\ldots&0&0& \cr
                         &\cr
                         \vdots&\vdots&&\ddots&\ddots&\vdots   \cr
                         0&0&\ldots&0&\sqrt{[F-1]}&0& } \cr
                      ~~~~~~ \cr
                 \pmatrix{0&0&0&\ldots&0&0& \cr
                         \sqrt{1(F-1)}&0&0&\ldots&0&0& \cr
                         0& \sqrt{2(F-2)}&0&\ldots&0&0& \cr
                         &\cr
                         \vdots&\vdots&&\ddots&\ddots&\vdots   \cr
                         0&0&\ldots&0&\sqrt{(F-1)1}&0& } 
\end{array} 
\right.
Q^+_{1-{1\over F}}=\left\{
\begin{array}{ll} 
                  \pmatrix{0&0&0&\ldots&0&\left\{\sqrt{[F-1]!}\right\}^{-1} \cr
                          0&0&0&\ldots&0&0& \cr
                          0&0&0&\ldots&0&0& \cr
                          &\cr\
                          \vdots&\vdots&&&\ddots&\vdots& \cr 
                           0&0&0&\ldots&0&0&} \cr
                        ~~~~~~ \cr
                \pmatrix{0&0&0&\ldots&0&1/(F-1)!\cr
                          0&0&0&\ldots&0&0& \cr
                          0&0&0&\ldots&0&0& \cr
                          &\cr\
                          \vdots&\vdots&&&\ddots&\vdots& \cr 
                           0&0&0&\ldots&0&0&} 
  
\end{array} 
\right.
\end{array}
\eeqa  
}

\noindent
with $[a] = {q^{-a/2}-q^{a/2} \over q^{-1/2} - q^{1/2}}$, 
$[F-1]!= [F-1] [F-2] \cdots [2] [1]$ and $q=\exp{(2i\pi /F)}$.
Of course the three sets of matrices given in (\ref{eq:q}) and
(\ref{eq:q1}) are related by a conjugation transformation (or a rescalling
of the vectors which belong to the representation --see after--).

From the basic conjugation we obtain immediately the associated representation
for the $Q^-$ charges
\beqa
\label{eq:qdag}
Q^-_{-{1\over F}}&=&\left(Q^+_{-{1\over F}}\right)^\dag \\
Q^-_{1-{1\over F}}&=&\left(Q^+_{1-{1\over F}}\right)^\dag  \nonumber           
\eeqa  

\noindent
There are two consequences of the exhibited representations.
\begin{enumerate}
\item A direct calculation shows that the two charges $Q^+_{-1/F}$
and $Q^-_{-1/F}$  satisfy quadratic relations. 
\begin{enumerate}
\item
In the case of the
first series we obtain  the $q-$oscil\-lator algebra introduced by 
Biedenharn and Macfarlane \cite{bm}

\beqa
\label{eq:qos}
&&Q^-_{-1/F} Q^+_{-1/F} - q^{\pm 1/2} Q^+_{-1/F} Q^-_{-1/F} = q^{\mp N/2} \\
\nonumber
&&[N,Q^+_{-1/F}]=Q^+_{-1/F} \\
&&[N,Q^-_{-1/F}]=-Q^-_{-1/F},  \nonumber
\eeqa
\noindent
with $N={\mathrm{diag}}(0,1,\cdots,F-1)$ the number operator 
(which can be expressed with $Q_{-1/F}^\pm$).
\item
For the second choice we have
\beqa
\label{eq:sl}
&&[Q^-_{-1/F}, Q^+_{-1/F}]  =  N =\mathrm{diag} (F-1,F-3,\cdots,
1-F) \\
&&[ N,Q^\pm_{-1/F}]= \mp 2  Q^\pm_{-1/F}, \nonumber
\eeqa
showing that the $Q$ generate the $F-$dimensional representation of
$sl(2,\RR)$. 
\end{enumerate}
Among those two matrix representation of the FSUSY algebra (and eventually
others) we were not able to find arguments to select one rather the other
{\it i.e.} to obtain  {\it naturally} and independently of
{\it any} matrix realization  a  quadratic relation among $Q^+_{-1/F}$ and
$Q^-_{-1/F}$ which characterizes the structure of the FSUSY algebra. 
Some indications in this direction should be  given. We can first notice
the property that the usual   superspace construction of SUSY,  by the
help of Grassmann variables,  can be 
generalized,  and an adapted version has already been built
within the framework of FSUSY, at least when $D=1,2$ 
\cite{fr,prs,fsusy1d,fsusy2d}. Secondly, we can observe that 
the quantization of the algebra generated by $Q^+_{-1/F}$
and its conjugate $Q^-_{-1/F}$ (variables fullfiling 
$\theta^F=0$ and generalizing the well-known Grassmann variables)
is related with the $q-$deformed 
Heisenberg algebra \cite{hq}.  In other words we might have relations like
$Q^+_{-1/F} \sim \theta$, $Q^-_{-1/F} \sim \partial_\theta$ and 
$\partial_\theta \theta -q \theta \partial_\theta \sim 1$.
Furthermore it is known that the algebra generated by $\theta$ and
$\partial_\theta$ is equivalent to the $q-$oscillators \cite{bm}. 
These two  remarks  are surely related  and
can be compared with the fact that the quantization of
the Grassmann algebra is the Clifford algebra.

As a consequence, the representation
built with the $Q$'s is unitary.
Indeed, the quadratic relations  (\ref{eq:qos}) or (\ref{eq:sl}) enable us 
to prove that the norm of the vector $\left(Q^+_{-1/F}\right)^n$ $ |0>$, 
with $n=0,\cdots,F-1$ and $|0>$
the primitive vector on which the representation span by $Q^{\pm}_{-1/F}$
is built, is positive. This result can be obtained even more simply,
using the results of the $q-$oscillators for the first series \cite{bm},
or by proving that the matrices given in (\ref{eq:sl}) can be mapped to
the $F \times F$ hermician matrices of $SU(2)$, which generate unitary 
representation (see after).
The deep reason for the emergence of a quadratic  structure is 
the non-faithfulness of the representation.
Indeed, relations (\ref{eq:PQr}) are
not strong enough to order the monomials in such a way that, say
$Q^+_{-1/F}$, is always on the left of $Q^+_{1-1/F}$, and the number of
monomials increase with their degree. If we have  a finite-dimensional
representation then it means that we have obtained  quadratic relations:
this allows us to order the monomials.
\item 
We can observe directly  that
$$Q^+_{1-1/F}= {1 \over [F-1]!} \left(Q^-_{-1/F}\right)^{F-1};$$
for the first choice, and
$$Q^+_{1-1/F}={1 \over ((F-1)!)^2} \left(Q^-_{-1/F}\right)^{F-1}$$
for the second.
Because of this constraint, the  $Q_{-1/F}^\pm$ alone span the representation
of the FSUSY algebra.
\end{enumerate}

We note that the representations   built with
the matrices $Q_{-{1 \over F}}$ and $Q_{1-{1 \over F}}$ can be obtained
in a way similar  to the way one obtains representations of  SUSY \cite{fs}.
We start with a vacuum $\Omega_\lambda$ in the spin$-\lambda$
representation
of $SO(1,2)$. On-shell, using the results established in \cite{jn,p},
we have the following decomposition 
$$\Omega_\lambda = \Omega_{h=\lambda}^+ \oplus \Omega_{h=-\lambda}^-,$$
with  two states of helicity $\pm \lambda$ and  positive/negative energy.
These two vacua are $CP-$conj\-uga\-te and allow us to build a
$CP-$invariant
representation. This constraint of $CP$ invariance is very strong, because
as soon as we have chosen the representation built from
$\Omega_{h=\lambda,+}$,
the  one built from  $\Omega_{h=-\lambda,-}$ is not arbitrary.
Altogether, with (\ref{eq:q1}) and (\ref{eq:qdag})  we get the representation
($Q^-_{-1/F} \Omega_{h=\lambda}^+=0, Q^+_{-1/F} \Omega_{h=-\lambda}^-=0$,
and for our normalization we have chosen the first choice for the $Q$'s)

$$\vbox{\offinterlineskip \halign{
\tv# & \cc{#} & \tv# & \cc{#}  & \tv# &
\cc{#} & \tv# & \cc{#} & \tv# & \cc{#}& \tv# \cr
\noalign{\hrule}
&\cc{states}&&\cc{helicity} &&\cc{states}&&\cc{helicity} & \cr
\noalign{\hrule}
&$\Omega_{\lambda}^+$&&$\lambda$&
&$\Omega_{-\lambda}^-$&&$-\lambda$& \cr
\noalign{\hrule}
&$Q^+_{-1/F}\Omega_{\lambda}^+$&&$\lambda-1/F$&
&$Q^-_{-1/F}\Omega_{-\lambda}^-$&
&$-\lambda+1/F$& \cr
\noalign{\hrule}
&$\vdots$&& && &&$\vdots$& \cr
\noalign{\hrule}
&${\left(Q^+_{-1/F}\right)^a \over \sqrt{[a]!}}\Omega_{\lambda}^+$&
&$\lambda-a/F$&
&${\left(Q^-_{-1/F}\right)^a \over \sqrt{[a]!}}\Omega_{-\lambda}^-$&
&$-\lambda+a/F$ &\cr
\noalign{\hrule}
&$\vdots$&& && &&$\vdots$& \cr
\noalign{\hrule}
&${\left(Q^+_{-1/F}\right)^{F-1} \over \sqrt{[F-1]!}}\Omega_{\lambda}^+$&
&$\lambda-(F-1)/F$&
&${\left(Q^-_{-1/F}\right)^{F-1}\over \sqrt{[F-1]!}}\Omega_{-\lambda}^-$&
&$-\lambda+(F-1)/F$ &\cr
\noalign{\hrule}
}}$$

\noindent
The states of positive energy and helicity ($\lambda,\lambda -{1 \over F},
\dots,\lambda -{F-1 \over F}$) are   $CP-$ conjugate to the states 
of negative  negative energy  and helicity ($-\lambda,-\lambda +{1 \over
F}, \dots,-\lambda +{F-1 \over F}$), and following the remarks given here above
it is known that the representation  is unitary.
An interesting consequence of the second choice for the $Q-$matrices is the
fact that the representation of the FSUSY algebra belong to a $F-$dimensional
representation of $SU(2)$. Indeed, it is easy to check that the matrices
$K_1=1/2\left(Q^+_{-1/F}+Q^-_{-1/F}\right),
 K_2=i/2\left(Q^+_{-1/F}-Q^-_{-1/F}\right)$ and $K_3=N/2$ are unitary
and generate the $SU(2)$ algebra.

Hence, FSUSY is a direct generalization of SUSY in the sense that these
fractional spin states or anyons are connected by FSUSY transformations. 
The next step
would be to construct explicitly a Lagrangian invariant under a FSUSY 
transformation which mixes these states, as  has been done in one and
two dimensions \cite{fsusy,am,fr,prs,fsusy1d,fsusy2d}. As a starting point,
one could use the lagrangian formulation of anyonic fields given 
in \cite{jn,p}.

To conclude this general study of the algebra,  it is
of great interest to mention some  properties when $F$ is not a prime
number.
Assuming $F=F_1 F_2$,  we have $F_1SP_{1,2} 
\subset FSP_{1,2}$. This property was already observed
in two dimensions in the second paper of \cite{prs}. So, this inclusion
(which can also be proven in one dimension) is a general property of FSUSY
and does not depend on the dimension. 
To prove this statement, we focus on the case where we have only
the $Q^+$ charges and we omit the $+$ superscript. 

If we define $\left(Q_{-{1\over F}} \right)^{F_2}=Q_{-{1\over F_1}} $, 
using the algebra
we can build, from the spin$-{1 \over F}$ representation, a 
spin$-{1 \over F_1}$ representation
of $SO(1,2)$ : $Q_{n-{1\over F_1}} \sim \left[J_+,\dots,
\left[J_+, Q_{-{1\over F_1}},\right],\dots\right]$ where $J_+$ has
been applied $n-$times. Using the Jacobi  identities (\ref{eq:J}),
we can construct an algebraic generalization of (\ref{eq:PQ}) which
mixes the spin$-{1\over F}$ and spin$-{1\over F_1}$ anyonic operators.

The case where $F$ is an even number is special because the spin$-1/2$ 
representation is finite, so we have the same constraints as before for
(\ref{eq:PQ}).
From these inclusions of algebras, we are able to build  sub-algebras with 
smaller symmetries  when $F$ is not a
prime number. In such a situation, the
$F-$multiplet  of $FSP_{1,2}$  splits into
$F_2$~~$F_1-$multiplets of  $F_1SP_{1,2}$
$$\Phi_\lambda^{(F)} = \bigoplus \limits_{a=0}^{F_2-1} 
\Phi_{\lambda + {a\over F}}^{(F_1)}.$$
The $F_1-$multiplet $\Phi_{\lambda + {a\over F}}^{(F_1)}$ is built from the
vacuum $\Omega_{\lambda + {a\over F}}$. This can be checked directly from
the
definitions and  using the representations ) or the matrices  (\ref{eq:q1})
and (\ref{eq:q}).

In this letter, we have explicitly constructed  non-trivial algebraic
extensions
of the $3D$ Poincar\'e algebra that go beyond the supersymmetric ones. 
The study of their representations enables us to show that these symmetries
connect the fractional spin states given in (17-18).
 We have pointed out an interesting
classification of these algebras by means of  the decomposition of $F$ 
(the order  of FSUSY)  as a  product of prime numbers. 
This leads to sub-systems with
smaller symmetries. A first application of these algebras, would be to
build a Lagrangian formulation where  FSUSY, among anyonic fields, is
manifest.
This could lead to some generalizations of the well known Wess-Zumino model
\cite{wz}. A further application would be to gauge FSUSY along the lines
given 
in \cite{fr}, after having studied the massless representations of the
algebra (\ref{eq:P}),(\ref{eq:Q}) and (\ref{eq:PQ}).

Recently, a very interesting interpretation of supersymmetry and fractional
supersymmetry in one dimension was given as an appropriate limit of the 
braided line \cite{bl}. Is it possible to understand, along these lines,
how
supersymmetry and fractional supersymmetry emerge in two and three
dimensions
and to prove that when the dimension is higher than three
 only SUSY is allowed ?

Finally, it  should  be  interesting  to understand the consequences 
of  the FSUSY extensions of the Poincar\'e algebra, in relation with three
dimensional physics.
\vskip.5truecm 
We would like to thank A. Comtet, E. Dudas, M. Plyushchay,
Ph. Revoy and C. A. Savoy
for  critical remarks and useful discussions.
We would also like to thank the referee for his remarks and suggestions.\\


\end{document}